\documentclass[12pt]{article}
\usepackage{html}
\usepackage{epsf}
\usepackage{graphicx}
\usepackage{amssymb}
\begin{document}
\title{MATTERS OF GRAVITY, The newsletter of the APS Topical Group on 
Gravitation}
\begin{center}
{ \Large {\bf MATTERS OF GRAVITY}}\\ 
\bigskip
\hrule
\medskip
{The newsletter of the Topical Group on Gravitation of the American Physical 
Society}\\
\medskip
{\bf Number 34 \hfill Fall 2009}
\end{center}
\begin{flushleft}
\tableofcontents
\vfill\eject
\section*{\noindent  Editor\hfill}
David Garfinkle\\
\smallskip
Department of Physics
Oakland University
Rochester, MI 48309\\
Phone: (248) 370-3411\\
Internet: 
\htmladdnormallink{\protect {\tt{garfinkl-at-oakland.edu}}}
{mailto:garfinkl@oakland.edu}\\
WWW: \htmladdnormallink
{\protect {\tt{http://www.oakland.edu/physics/physics\textunderscore people/faculty/Garfinkle.htm}}}
{http://www.oakland.edu/physics/physics_people/faculty/Garfinkle.htm}\\

\section*{\noindent  Associate Editor\hfill}
Greg Comer\\
\smallskip
Department of Physics and Center for Fluids at All Scales,\\
St. Louis University,
St. Louis, MO 63103\\
Phone: (314) 977-8432\\
Internet:
\htmladdnormallink{\protect {\tt{comergl-at-slu.edu}}}
{mailto:comergl@slu.edu}\\
WWW: \htmladdnormallink{\protect {\tt{http://www.slu.edu/colleges/AS/physics/profs/comer.html}}}
{http://www.slu.edu//colleges/AS/physics/profs/comer.html}\\
\bigskip
\hfill ISSN: 1527-3431

\bigskip

DISCLAIMER: The opinions expressed in the articles of this newsletter represent
the views of the authors and are not necessarily the views of APS.
The articles in this newsletter are not peer reviewed.

\begin{rawhtml}
<P>
<BR><HR><P>
\end{rawhtml}
%{\bf \Large Contents:}
\end{flushleft}
\pagebreak
\section*{Editorial}

The next newsletter is due February 1st.  This and all subsequent
issues will be available on the web at
\htmladdnormallink 
{\protect {\tt {http://www.oakland.edu/physics/Gravity.htm}}}
{http://www.oakland.edu/physics/Gravity.htm} 
All issues before number {\bf 28} are available at
\htmladdnormallink {\protect {\tt {http://www.phys.lsu.edu/mog}}}
{http://www.phys.lsu.edu/mog}

Any ideas for topics
that should be covered by the newsletter, should be emailed to me, or 
Greg Comer, or
the relevant correspondent.  Any comments/questions/complaints
about the newsletter should be emailed to me.

A hardcopy of the newsletter is distributed free of charge to the
members of the APS Topical Group on Gravitation upon request (the
default distribution form is via the web) to the secretary of the
Topical Group.  It is considered a lack of etiquette to ask me to mail
you hard copies of the newsletter unless you have exhausted all your
resources to get your copy otherwise.

\hfill David Garfinkle 

\bigbreak

\vspace{-0.8cm}
\parskip=0pt
\section*{Correspondents of Matters of Gravity}
\begin{itemize}
\setlength{\itemsep}{-5pt}
\setlength{\parsep}{0pt}
\item John Friedman and Kip Thorne: Relativistic Astrophysics,
\item Bei-Lok Hu: Quantum Cosmology and Related Topics
\item Veronika Hubeny: String Theory
\item Beverly Berger: News from NSF
\item Luis Lehner: Numerical Relativity
\item Jim Isenberg: Mathematical Relativity
\item Lee Smolin: Quantum Gravity
\item Cliff Will: Confrontation of Theory with Experiment
\item Peter Bender: Space Experiments
\item Jens Gundlach: Laboratory Experiments
\item Warren Johnson: Resonant Mass Gravitational Wave Detectors
\item David Shoemaker: LIGO Project
\item Stan Whitcomb: Gravitational Wave detection
\item Peter Saulson and Jorge Pullin: former editors, correspondents at large.
\end{itemize}
\section*{Topical Group in Gravitation (GGR) Authorities}
Chair: Stan Whitcomb; Chair-Elect: 
Steve Detweiler; Vice-Chair: Patrick Brady. 
Secretary-Treasurer: Gabriela Gonzalez; Past Chair:  David Garfinkle;
Delegates:
Lee Lindblom, Eric Poisson,
Frans Pretorius, Larry Ford,
Scott Hughes, Bernard Whiting.
\parskip=10pt

\vfill
\eject

\section*{\centerline
{New MOG correspondents}}
\addtocontents{toc}{\protect\medskip}
\addtocontents{toc}{\bf GGR News:}
\addcontentsline{toc}{subsubsection}{
\it New Matters of Gravity Correspondents, by David Garfinkle}
\parskip=3pt
\begin{center}
David Garfinkle, Oakland University
\htmladdnormallink{garfinkl-at-oakland.edu}
{mailto:garfinkl@oakland.edu}
\end{center}

MOG correspondent is now a rotating office.  This and the next few issues
will usher in new correspondents.  I want to thank the correspondents for 
their long and faithful service to Matters of Gravity.  I also want to 
welcome the new correspondents: Veronika Hubeny in string theory, Luis Lehner
in numerical relativity, and Jim Isenberg in mathematical relativity.

\bigskip
\section*{\centerline
{we hear that \dots}}
\addtocontents{toc}{\protect\medskip}
\addcontentsline{toc}{subsubsection}{
\it we hear that \dots , by David Garfinkle}
\parskip=3pt
\begin{center}
David Garfinkle, Oakland University
\htmladdnormallink{garfinkl-at-oakland.edu}
{mailto:garfinkl@oakland.edu}
\end{center}

Patrick Brady has been elected Vice-Chair of GGR.  Scott Hughes and 
Bernard Whiting have been elected members at large of the GGR Executive
Committee.

Hearty Congratulations!

\vfill\eject

\section*{\centerline
{Numerical Relativity Studies of Black Hole Kicks}}
\addtocontents{toc}{\protect\medskip}
\addtocontents{toc}{\bf Research briefs:}
\addcontentsline{toc}{subsubsection}{
\it Numerical Relativity Studies of Black Hole Kicks, 
by Pablo Laguna}
\parskip=3pt
\begin{center}
Pablo Laguna, Georgia Institute of Technology
\htmladdnormallink{plaguna-at-gatech.edu}
{mailto:plaguna@gatech.edu}
\end{center}
\newcommand{\KMS}{\mbox{km s}^{-1}\,}  

Halfway through the present decade, numerical relativity experienced a
revolution that triggered a boom in modeling binary black holes as
never seen before. Work through the 70's, 80's and 90's paved the way
for the four seminal papers responsible for this revolution: one on
the first binary black hole orbit~\cite{Bruegmann:2003aw} and three on
the first inspiral and
merger~\cite{2005PhRvL..95l1101P,2006PhRvL..96k1101C,2006PhRvL..96k1102B}.

The numerical relativity community immediately realized that these
breakthroughs opened the door to investigations of the gravitational
recoil experienced by the final black hole in binary mergers,
investigations in a regime only accessible to the numerical relativity
machinery: the fully non-linear phase of late inspiral and plunge. It
was well known that a net flux of linear momentum is emitted during
the merging of the binary~\cite{PhysRev.128.2471,1973ApJ...183..657B}
and that as a consequence the remnant black hole will experience a
kick. What was unclear was the particulars of the linear momentum
emission and its contribution to the kicks during the last few orbits
and merger. What was also uncertain was the accuracy of post-Newtonian
approximations in estimating
kicks~\cite{1983MNRAS.203.1049F,1984MNRAS.211..933F,
  1992PhRvD..46.1517W,2005ApJ...635..508B,2006PhRvD..73l4006D}. 

In the span of almost three years, numerical relativity efforts around
the world produced a plethora of results on black hole kicks, results
that not only helped us gain insight on Einstein's theory of general
relativity but also delivered surprises. One of these surprises was
the kick magnitudes of a few thousands km/s, triggering a tremendous
excitement in the astrophysics community because of their implications
for the growth of massive black holes, galaxy formation scenarios and
potential electromagnetic signatures.

This research brief attempts to summarize highlights from the
numerical relativity studies of gravitational recoil. Specific details
of the work can be found in the references. In regard to some of the
astrophysical consequences of black hole kicks, the reader is
encourage to consult the following
references:~\cite{2008ApJ...689L..89K,2008ApJ...682..758S,2008ApJ...678L..81K,
  2008MNRAS.390.1311B,2009MNRAS.394..633D,2008ApJ...683L..21K}.

The first numerical relativity studies on gravitational recoil focused
on non-spinning, un-equal mass black hole binaries. The first kick
estimate that used the machinery of numerical relativity was obtained
by~\cite{2005CQGra..22S.387C} using the Lazarus approach. The first
study that took advantage of the breakthroughs in numerical relativity,
i.e. the moving puncture
methodology~\cite{2006PhRvL..96k1101C,2006PhRvL..96k1102B}, was
carried out by~\cite{2007CQGra..24...33H}. This work consisted mostly
of plunges and unfortunately had limitations in accuracy. Next, the
work by~\cite{2006ApJ...653L..93B} considered the case of a binary
with a 1.5:1 mass ratio, obtaining kicks between 86 and 97 km/s. A
much more comprehensive study followed these works, done
by~\cite{2007PhRvL..98i1101G}. This work expanded mass ratios $q =
m_2/m_1$ in the range $0.25 \le q \le 1$ or equivalently $0.16 \le
\eta \le 0.25$ with $\eta =q/(1+q)^2$. A maximum kick velocity of
$\sim 175$ km/s was obtained at $\eta \sim 0.2 \,(q \sim 0.38)$.

An interesting study aimed at unveiling the \emph{anatomy} of the
binary black hole recoil using a multipolar analysis was carried out
by~\cite{2008PhRvD..77d4031S}. One of the findings in this study was
that only the multipole moments up to and including $l=4$ are needed
to accurately reproduce within 2\% numerical relativity recoil
estimates. The simulations of the most extreme mass ratios of
initially non-spinning black-hole binaries were performed
by~\cite{2009PhRvD..79l4006G}. They considered a binary with 10:1
mass ratio ($q = 0.1$ and $\eta = 0.0826$) that completes three orbits
prior to merger, radiating $(0.415\pm0.017)\%$ of the total energy and
$(12.48\pm0.62)\%$ of the initial angular momentum in the form of
gravitational waves. The kick inflicted on the final black hole from
this merger is $(66.7\pm3.3)$km/s.

The attention then shifted to gravitational recoil from the merger of
spinning black hole binaries. The first study of this kind was done
by~\cite{2007ApJ...661..430H}. This study focused on equal-mass black
holes with anti-aligned spins along the orbital angular momentum
direction. The binary mergers produced a kick on the final black hole
of $\sim 475\,a/M$ km/s with $a$ the spin parameter of the black hole.
Kicks of this magnitude could result in the ejection of black holes
from the core of a dwarf galaxy. Similar conclusions were found in
work by~\cite{2007PhRvL..99d1102K} and \cite{2007ApJ...668.1140B}.
Around the same time, kicks in the head-on collision of spinning black
holes were reported by~\cite{2007PhRvD..76j4026C}. A study
by~\cite{2007ApJ...659L...5C} considered a more generic
configuration: mass ratio of 2:1, with the smaller black hole
non-spinning and the larger, a rapidly spinning hole oriented $45^o$
with respect to the orbital plane. The remnant black hole received a
kick of 454 km/s. A closer look at the components of this kick
velocity hinted that much larger kicks could be obtained if the black
holes have spins in the orbital plane and counter-aligned. The first
example of these super-kick velocities ($\sim 1300$ km/s) was reported
by Sperhake and collaborators~\cite{2007PhRvL..98w1101G} at the
workshop on the interface between post-Newtonian theory and numerical
relativity in February, 2007 at Washington University in St. Louis.
Results from a similar simulation were also included in the final
version of Ref.~\cite{2007ApJ...659L...5C}. Followup work
by~\cite{2007PhRvL..98w1102C} showed that kicks of up to $\sim 4000$
km/s can be obtained for maximally-rotating holes for these
in-plane/counter-aligned configurations. Finally,
\cite{2008PhRvD..78b4039D} evolved black hole binaries with the
highest spins simulated thus far ($a/m_h \sim 0.925 $) in a set of
configurations with the spins counter-aligned and pointing in the
orbital plane, which maximizes the recoil velocities of the merger
remnant (i.e. super-kick configuration).

The foundations were in place for a series of studies that effectively
explored the parameter space. Work by~\cite{2007PhRvD..76h4032H}
analyzed the spin dynamics of the individual black holes in connection
with the gravitational recoil. \cite{2007PhRvD..76f1502T} carried out
the first relatively large set of simulations of equal mass black
holes with general spin orientations. \cite{2007PhRvD..76l4002P}
performed a systematic investigation of spin-orbit aligned
configurations and introduced a phenomenological expression for the
recoil velocity as a function of spin ratio.
\cite{2008PhRvD..77l4047B} showed that the recoil is mostly given by
the difference between the $(l=2,m=\pm2)$ modes of $\Psi_4$ and that
the dominant part of this contribution takes place within $30M$ before
and after the merger. \cite{2008ApJ...679.1422R} constructed ``spin
diagrams'' that allow one to estimate the recoil velocity and spin of
the remnant black hole in terms of the spins of the merging black
holes. \cite{2008ApJ...682L..29B} showed that kicks perpendicular to
the orbital plane scale as $\sim\eta^3$. A scaling such as this would
tend to suppress kicks for binary mergers with large mass ratios.
Recent work by~\cite{2009PhRvD..79f4018L} indicates that the scaling
seems to be instead $\sim\eta^2$, in agreement with post-Newtonian
scalings. \cite{2007PhRvD..76h1501K} proposed a quasi-local formula
for the linear momentum of black-hole horizons inspired by the
formalism of quasi-local horizons.

An interesting outcome from all of these studies has been a
parametrized empirical formula to estimate kicks. The formula was
introduced by \cite{2007ApJ...659L...5C} motivated by post-Newtonian
scalings:
\begin{equation}
\vec{V}_{\rm recoil}(q,\vec\alpha_i)=v_m\,\hat{e}_1+
v_\perp(\cos(\xi)\,\hat{e}_1+\sin(\xi)\,\hat{e}_2)+v_\|\,\hat{e}_z,
\end{equation}
with
\begin{equation}\label{eq:vm}
v_m=A\frac{q^2(1-q)}{(1+q)^5}\left(1+B\,\frac{q}{(1+q)^2}\right),
\end{equation}
\begin{equation}\label{eq:vperp}
v_\perp=H\frac{q^2}{(1+q)^5}\left(\alpha_2^\|-q\alpha_1^\|\right),
\end{equation}
\begin{equation}\label{eq:vpar}
v_\|=K\cos(\Theta-\Theta_0)\frac{q^2}{(1+q)^5}\left|\vec\alpha_2^\perp-q\vec\alpha_1^\perp\right|,
\end{equation}
with $\vec{\alpha}_i=\vec{S}_i/m_i^2$, $\vec S_i$ and $m_i$ the spin
and mass of holes, $\hat{e}_1$ and $\hat{e}_2$ orthogonal unit vectors
in the orbital plane, $\xi$ a measure of the angle between the unequal
mass and spin kick contributions and $\Theta$ the angle between the
in-plane component of $\vec \Delta\equiv (m_1+m_2)({\vec S_2}/m_2
-{\vec S_1}/m_1)$ and the infall direction at merger. The dependences
and parameters in this formula have been verified and fixed by
numerical relativity simulations: $A = 1.2\times 10^{4}\ \KMS$, $B =
-0.93$, $H = (6.9\pm0.5)\times 10^{3}\ \KMS$ and $K=(6.0\pm0.1)\times
10^4\ \KMS$. Details on the \emph{history} of this empirical formula
can be found in the the paper by \cite{2008PhRvD..77d4028L} where a
systematic study is presented showing the accuracy of this heuristic
model. The empirical kick formula has become a great resource in
astrophysical studies of black hole mergers and their remnants since
it provides a tool to estimate kicks without the need of numerical
relativity simulations.

Finally, just recently, \cite{2009PhRvL.102d1101H} investigated the
gravitational recoil produced in hyperbolic encounters. The encounters
were designed to be plunge dominated. The motivation was to avoid as
much as possible the averaging of the beamed linear momentum radiation
as the black holes inspiral. As a consequence, these encounters
produced kick velocities as large as 10,000 km/s. It is very unlikely,
however, that these type of extreme scattering mergers have
significant astrophysical relevance.

The gravitational recoil of the remnant from black hole mergers is a
beautiful example of the tremendous potential that numerical
relativity has as a tool of discovery. The simulations unveiled
unexpected results (e.g. super-kicks) that trigger tremendous
excitement in the astrophysical community. It is almost certain that
this was not a rare event. As we continue to explore other non-linear
gravitational phenomena, many more surprises are yet to come from
numerical relativity studies.

\vfill\eject

\section*{\centerline
{AbhayFest}}
\addtocontents{toc}{\protect\medskip}
\addtocontents{toc}{\bf Conference reports:}
\addcontentsline{toc}{subsubsection}{
\it AbhayFest, 
by Chris Beetle}
\parskip=3pt
\begin{center}
Chris Beetle, Florida Atlantic University 
\htmladdnormallink{cbeetle-at-physics.fau.edu}
{mailto:cbeetle@physics.fau.edu}
\end{center}
AbhayFest was held in State College, Pennsylvania from Thursday, June 4 through Saturday, June 6, 2009 to mark the sixtieth birthday of Abhay Ashtekar.  The conference was a great success, well and boisterously attended.  The organizing committee of Martin Bojowald, Jorge Pullin, Paul Sommers and Randi Neshteruk did a wonderful job putting together a diverse and stimulating array of talks reflecting Abhay's varied and passionate scientific interests.

The scientific program comprised morning sessions of forty-five minute plenary presentations on each of its three days, followed by afternoon sessions on the first two days of more focussed thirty minute talks. The entire proceedings are on the conference web site: 
\smallskip
\htmladdnormallink
{\protect {\tt{http://gravity.psu.edu/events/abhayfest/}}}
{http://gravity.psu.edu/events/abhayfest/}\\
\smallskip

Both the original slides and an audio recording are available for each presentation.

The first scientific talk on Thursday morning was from Jim Hartle, arguing that the viability of semi-classical physics at late times in a quantum gravitational universe dramatically restricts its initial quantum state in a way that could explain the origins of both the arrow of time and inflation.  This talk produced easily the most memorable quote from the conference when, as the audience continued to get its bearings after many weather-induced late-night and early-morning arrivals to State College, the speaker wondered aloud, ``I explained the whole universe and these are all the questions I get?''  It seemed the answer, to leading order at least, was in the affirmative.  Jim's talk was followed by Klaus Fredenhagen describing the mathematical physics of the analogue of non-normalizable eigenfunctions in the algebraic approach to quantum field theory and their relation to the problem of identifying time as an observable in quantum theory.  Interestingly, the approach predicts non-commutativity of spacetime, and even the time axis naturally becomes the Toeplitz quantization of the real line.  The final talk on Thursday morning was from Gary Horowitz on the phenomenological similarity between the analysis of quantum black-hole ``singularities'' in string theory and the corresponding Ashtekar--Bojowald analysis in loop quantum gravity.  It seems that both theories point clearly at a new paradigm that the quantum-mechanical structure of the deep interior of a black hole foils any effort to define a global event horizon even in the semi-classical limit.

Thursday's afternoon session featured talks by prot\'eg\'es of Abhay's.  The first was from Bernd Br\"ugmann, summarizing the recent dramatic advances in numerical relativity in general, and the emergence of the puncture method in particular.  This was followed by a presentation by Badri Krishnan summarizing the current state of, and potential scientific payoff from, searches for gravitational waves from neutron stars at LIGO.  Steve Fairhurst followed this with a related talk dealing with the search for gravitational waves created in binary coalescence.  The second half of the afternoon session was devoted to issues in mathematical and quantum general relativity.  It began with a presentation by Jon Engle summarizing the theory and application of isolated and dynamical horizons in classical and numerical gravity.  This was followed by Alex Corichi reporting on settled and open issues in the statistical analysis of black hole entropy in loop quantum gravity.  Finally, Madhavan Varadarajan explored how the Ashtekar--Bojowald picture of black holes, described in Gary Horowitz's earlier talk, sheds light on the Bekenstein information-loss paradox from the theory of quantum fields in curved (classical) spacetime.

Sir Roger Penrose offered a public lecture to an overflow crowd on Thursday night entitled ``Fashion, faith and fantasy: How big is infinity?"  Sadly, your intrepid reporter forgot to bring his notebook to dinner, whence he had to rush to the lecture.  However, Roger's lecture summarized a previous series of lectures he gave at Princeton in October 2003.  Video of these lectures may be found at 

\smallskip
\htmladdnormallink
{\protect {\tt{http://www.princeton.edu/WebMedia/lectures/}}}
{http://www.princeton.edu/WebMedia/lectures/}\\
\smallskip

and would presumably do a much better job of communicating the key ideas than any summary here.

Friday's morning session opened with a presentation from Carlo Rovelli provocatively titled ``What is a particle?''  It observed the tension between two distinct notions of particles in quantum field theory, one the global notion of particles as Poincar\'e covariant states in Fock space and the other the more local notion of a particle as a disturbance that causes a localized detector to respond.  It suggested a possible resolution to this tension based on a careful analysis of the relation between the two classes of states.  This was followed by a report from Bob Wald on the status of his recent work, mainly in collaboration with Stefan Hollands, on the algebraic approach to quantum field theory in curved spacetimes.  The central concepts in this work include a microlocal spectrum condition on states and an operator product expansion in the observable algebra, the existence of which play the roles in curved spacetime that Poincar\'e invariance and the uniqueness of the corresponding vacuum do in ordinary Fock-space models.  Finally, Thomas Thiemann gave a status report on loop quantum gravity.  The foundations of this theory have become much better understood in recent years and, though the final picture is not yet clear, uniqueness results have been found that place useful restrictions on the quantum dynamics of the theory and that clarify its semi-classical limit.

The afternoon session on Friday began with Laurent Friedel presenting an elegant analysis of the physical features of states of loop quantum gravity restricted to a particular graph based on a geometric quantization of the corresponding phase space of classical holonomies.  This was followed by Jerzy Lewandowski describing an attempt to bridge canonical loop quantum gravity to spin foam methods using a method that applies to general states of the theory, not just to spin-network eigenstates, without relying on Regge calculus.  Lee Smolin followed with a discussion of unimodular gravity, a theory in which the cosmological constant is a dynamical (rather than a kinematical) constant, and in which a natural conjugate momentum to the cosmological constant hints at a resolution to the problem of time.  The second half of the session began with a talk by Claes Uggla describing an ``asymptotically silent'' class of singularities whose physical structure can be profitably analyzed using conformal methods in classical general relativity.  Jim Isenberg followed with a presentation centered on an approach to finding initial data for the general relativistic $n$-body problem using surgery of 3-manifolds.  The final speaker on Friday afternoon was Naresh Dadhich, who described a general analysis of higher-order polynomial Lagrangians for relativistic gravity that is broad enough to include the Lovelock theory.

A conference banquet was held on Friday night during which many well-wishers, many attired nearly as nattily as the man himself, wished Abhay well.  The occasion was marked by exactly the appropriate degree of ardent purposefulness.  And there was free wine.  Audio recordings remain mercifully unavailable from the conference web site.

Your correspondent's notes from Saturday morning's session are slightly sketchier, a fact not unrelated to Friday's free wine and old friends.  Bernd Schmidt began with a presentation of ongoing work developing an approach of Ehlers to the Newtonian limit of general relativity.  This approach offers, in particular, a natural way to develop solutions to the constraint equations on initial data in general relativity.  Rodolfo Gambini followed with a discussion of the issue of time in quantum gravity based on a scheme that combines the ``evolving'' Dirac observables proposed by Rovelli, among others, with the approach via conditional probabilities of Page and Wootters.  One nice feature of this scheme is the emergence of quantum limitations on the quality of a physical clock.  Roger Penrose concluded the conference with a presentation on the status of the conformal cyclic cosmology, which holds that only conformal curvature is relevant in quantum gravity.  One consequence of this intriguing proposal is that an absence of conformal curvature in the early universe could explain the absence of observable white holes that one would naively expect to find generically in a kinematically time-symmetric theory. 

Interested readers are once again directed to the conference web site for more details on the various talks, the true theses of which have doubtless been done scant justice here.  Finally, once again, happy birthday Abhay!

\vfill\eject

\section*{\centerline
{Eastern Gravity Meeting}}
\addtocontents{toc}{\protect\medskip}
\addcontentsline{toc}{subsubsection}{
\it Eastern Gravity Meeting, 
by Jeff Winicour}
\parskip=3pt
\begin{center}
Jeff Winicour, University of Pittsburgh
\htmladdnormallink{winicour+-at-pitt.edu}
{mailto:winicour+@pitt.edu}
\end{center}

On June 15 and 16 I had the pleasure of participating in the 2009 Eastern
Gravity Meeting held at Rochester Institute of Technology (RIT). The
campus of RIT was moved in 1968 from central Rochester to an outlying
area where the impressive transition from a small teaching institute to a
major research university has taken place. It is a spacious modern
campus, with a hotel and restaurants conveniently located within walking
distance. Part of the ongoing development has been the establishment of
the Center for Computational Relativity and Gravitation (CCRG) with a
large faculty involved in black hole and neutron star physics. 

The group in Rochester is a nice addition to the traditional strongholds
of general relativity along the NY Thruway in upstate NY at Syracuse and
Cornell. There was good participation, notwithstanding the competition
from several other meeting around the same time. In all, there were 49
participants representing 14 institutes, most of them from the eastern
states but some from as far away as California. The two day
schedule was packed with a full schedule of 15 minute talks. A nice
feature of the meeting was a \$200 prize for the best student
presentation, which was sponsored by the APS Topical Group on
Gravitation. This attracted a healthy number of student talks of very
impressive quality. The program along with abstracts and slides is
available at 
\htmladdnormallink
{\protect {\tt{http://ccrg.rit.edu/~EGM2009/}}}
{http://ccrg.rit.edu/~EGM2009/}\\

I was lucky to give the first talk, when everyone was was still fresh, on
a disembodied formulation of the boundary data for Einstein's equations
in the spirit of the purely 3D version of Cauchy data in terms of metric
and extrinsic curvature. This was followed by Maria Babiuc-Hamilton's
(Marshal U.) presentation of progress in carrying out characteristic
extraction of waveforms at null infinity using Cauchy data from an
interior binary black hole inspiral.

Next came a talk by Bruno Mundin, a CCRG postdoc, who described an efficient way
to simulate relativistic binaries using a boson star model based upon the
conformally flat approximation. This was the first of several talks by RIT
faculty, postdocs and students which highlighted the rapid progress that the
group has made in becoming a center for computer simulation of strong gravity
astrophysics. Joshua Faber gave preliminary results of a project to simulate
accretion discs around merging and kicked black holes, which revealed how the
inclination of the final disc is related to the kick direction; Carlos Lousto
proposed a simple empirical formula to describe the final remnant mass, spin and
recoil velocity from the merger of two black holes with arbitrary mass ratio and
spins;  David Merrit discussed how observations of the precession of stellar
orbits close to the central black hole in the Milky Way by next generation
telescopes can be used to test the no-hair theorem by measuring the angular
momentum and quadrupole moment of the black hole; Yosef Zlochower showed how the
algebraic class of a numerically evolved spacetime could be obtained by solving
the quartic equation governing the principle null directions; John Whelan
described how the cross-correlation of gravitational wave data streams, which is
used in searching for stochastic signals and bursts, could also be applied to
detect quasi-periodic waves; Hiroyuki Nakano presented a perturbative treatment
of the radiation recoil in a binary black hole merger; Fabio Antonini (student)
presented post-Newtonian N-body simulations of the tidal disruption of binaries
orbiting close to a supermassive black hole; Marcelo Ponce (student) reviewed
mechanisms by which the gravitational waves from a binary black hole can
stimulate electromagnetic waves from a surrounding accretion disc; and David
Sarnoff (student) showed how the simulation of test particle orbits about a Kerr
black hole is an ideal entry problem for learning how to use GPU's as a cheap
source of enhanced computing power in numerical relativity.  

{From} Cornell, Larry Kidder discussed how the combination of an
effective-one-body model with the Cornell-Caltech simulations of black
hole mergers can lead to improved gravitational wave data analysis;
Matthew Duez presented simulations revealing how spin affects black
hole-neutron star mergers and the resulting accretion disc; Geoffrey
Lovelace's talk focused on using the Landau-Lifshitz pseudotensor to
explore the momentum flow between binary black holes and the surrounding
spacetime; Rob Owen presented a method for computing multipole moments on
black hole horizons and gave some numerical results for the ringdown
following a merger; Jolyon Bloomfield (student) described an effective 4D
action for a given braneworld model, with the goal of shedding light on
dark energy; Francois Foucart (student) presented simulations showing how
the stiffness of the equation of state affects neutron star-black hole
binaries and the formation of an accretion disc; and Abdul Hussein Mroue
(student) presented single black hole evolutions using a new spectral BSSN
code.

{From} Syracuse, there were three student talks, by Collin Capano, Larne
Pekowsky and Peter Zimmerman, presenting their work on the LIGO and NINJA
data analysis projects.

Several upstate New York colleges were represented. Parker Troischt
(Hartwick College) presented an application of a Lagrangian fluid
formalism of relativistic MHD to the propagation of wave modes in a black
hole spacetime; and Munawar Karim (St. John Fisher College) described the
gravitational radiation energy density that would be necessary  to
account for the anomalous expansion of the universe.

{From} Princeton, Frans Pretorius discussed the now notorious possibility
of creating black holes in particle collisions at the LHC. Frans
presented preliminary results  of a project to simulate Planck energy
soliton collisions with the goal of shedding light on this issue; and a
Princeton student, Hans Bantilan, described how the AdS/CFT
correspondence might allow simulations of 5D Anti-deSitter space to provide
insight into high energy collisions of heavy ions.

{From} Penn State, Tomas Liko told how partition functions can be
calculated in Euclidean quantum gravity; Gianluca Calcagni described
recent developments in Horava's theory of gravity; Andrew Randono
described  the manner in which the internal spin angular momentum of a
spinor field is encoded in the gravitational field at infinity; Artur
Tsobanjan (student) presented a technique to handle constraints in
canonical quantum gravity; and Edward Wilson-Ewing (student) showed how
Bianchi I space-times can be quantized in the framework of loop quantum
gravity.

{From} New England, talks by Andreas Ross and a student James Gilmore
(Yale) showed how post-Newtonian corrections can be calculated using
effective field theory methods from particle physics; and an MIT student,
Sarah Vigeland, described how bumps on a black hole affects particle
orbits.

J. Brian Pitts of Notre Dame presented a novel approach to the old question of a
geometric interpretation of gravitational  energy-momentum pseudotensors. And
farther away from the Kavli Institute in Santa Barbara, Marc Favata discussed
how the "nonlinear memory" in a binary black hole merger, which causes a net
displacement of test particles after the gravitational wave passes, might be
calculable  with current numerical simulations and might even be observable with
ground or space-based interferometers.

On the final afternoon it came time to vote on the best student presentation.
The judges (Favata, Kidder, Pretorius and myself) were unanimous in the opinion
that all the students had done a great job, in terms of both the content and
presentation of their talks. We decided that the prize should be shared by
co-winners: Hans Bantilan for "Simulations of the Gravity Dual in an AdS/CFT
Correspondence" and Edward Wilson-Ewing for "Loop Quantum Cosmology of Bianchi I
Models". We were not sure whether the APS might apply stimulus funds to give
each the full prize but this dual award also led to the single glitch in the
meeting, which was otherwise perfectly orchestrated by the organizing committee
chair John Whelan. Even in this region where apples are named after the towns,
only one apple was available for the traditional local prize. How did the CCRG
director Manuela Campanelli resolve this dilemma? Check the website.

\vfill\eject

\section*{\centerline
{Benasque Workshop on Gravity}}
\addtocontents{toc}{\protect\medskip}
\addcontentsline{toc}{subsubsection}{
\it Benasque Workshop on Gravity, 
by Roberto Emparan and Veronika Hubeny}
\parskip=3pt
\begin{center}
Roberto Emparan , ICREA \& Universitat de Barcelona
\htmladdnormallink{emparan-at-ub.edu}
{mailto:emparan@ub.edu}
\end{center}
\begin{center}
Veronika Hubeny, Durham University 
\htmladdnormallink{veronika.hubeny-at-durham.ac.uk}
{mailto:veronika.hubeny@durham.ac.uk}
\end{center}

For the last fifteen years physicists have been gathering in Benasque, a
charming little town in the heart of the Spanish Pyrenees, to discuss
the most recent advances in their field of work. The idea of a Center
for Science (initially only physics, and since 1998 covering most fields
of knowledge) in the style of the Aspen Center for Physics was largely
the initiative of Spanish physicist Pedro Pascual (1934-2006), in whose
memory the Center has been recently renamed. The Center came of age this
Summer with the opening of its new building, a modern facility designed
as a place for work and interaction among researchers --- plenty of desk
space, omnipresent blackboards, common areas with coffee machines, and
multifunctional conference rooms.

It was in this venue that the workshop on ``Gravity: New perspectives
from strings and higher dimensions" took place during July 12-24, 2009,
bringing more than fifty participants to discuss the remarkable recent
progress in extending the field of application of General Relativity and
in understanding various aspects of higher-dimensional gravity --- as
well as taking the occasional opportunity for marvellous hiking that the
mountains around Benasque provide.

The scheduled program for the workshop was light, with two hours of
talks per day, plus a number of impromptu discussion sessions during
some of the evenings. The broad underlying theme was that General
Relativity, much like quantum field theory, can nowadays be regarded as
a tool to be applied to a wide variety of problems far beyond its
traditional realm of astrophysics and cosmology. Thus, some of the talks
discussed explicit applications of gravity to fields such as fluid
dynamics or superconductivity, while others focused on developing our
basic understanding of this tool, in particular in higher dimensions.

Three broad themes could be discerned among the subjects of the talks:
classification of higher-dimensional black hole spacetimes; methods for
constructing and analyzing new solutions; and the application of gravity
to learn about quantum-field-theoretical problems, or vice-versa. The
boundaries were of course not sharp and several of the talks could fit
into two or possibly all of these categories.

The classification of higher-dimensional spacetimes was the main focus
of talks by Harvey Reall, Stefan Hollands and Troels Harmark. Harvey
Reall opened the workshop with a presentation of his ongoing work on the
higher-dimensional extension of the Petrov classification. In four
dimensions this classification played an important role in the discovery
of a number of exact solutions (most notably, the Kerr black hole, as
well as the most general type-D metric of Plebanski and Demianski), and
it may be expected that a higher-dimensional generalization should be
equally useful. 

Stefan Hollands discussed the classification of five-dimensional black
hole solutions. In recent years a wealth of new exact solutions have
been discovered and powerful solution-generating techniques based on
inverse scattering methods have been applied to spacetimes with two
commuting spatial Killing fields (besides the stationarity timelike
isometry). Hollands and Yazadjiev have produced uniqueness theorems that
allow a complete characterization of black holes in this class. The
required data include the asymptotic conserved charges (mass and angular
momenta) as well as the `rod structure' introduced earlier by Emparan
and Reall and by Harmark for solutions with this symmetry.

Troels Harmark discussed how this scheme may be generalized to higher
dimensions, while overcoming some of its drawbacks. The `domain
structures', which characterize, up to volume-preserving
diffeomorphisms, the fixed-point sets of isometries of the spacetime
provide a simple and convenient way to distinguish among black hole
solutions.

The construction and analysis of new solutions for higher-dimensional
black holes, in exact or approximate manner, has seen dramatic progress
in recent years, and was the subject of talks by Pau Figueras, Toby
Wiseman and Niels Obers. We have already mentioned the application of
inverse-scattering techniques (in particular the Belinski-Zakharov
method) to five-dimensional vacuum gravity with three commuting abelian
isometries: this topic was reviewed by Figueras. 

There is a widespread appreciation that many of the most interesting
problems in modern General Relativity are unlikely to yield to analytic
techniques and that therefore numerical approaches are in order. Toby
Wiseman described his new approach to static numerical relativity (in
collaboration with Headrick and Kitchen) based on ideas borrowed from
Ricci-flow to turn Einstein's equations into a strictly elliptic
problem. Numerical studies were also discussed in an impromptu
presentation by Roberto Emparan of recent results on instabilities and
novel `pinched' phases of higher-dimensional rotating black holes (based
on work with Dias, Figueras, Monteiro and Santos that was posted in
arXiv during the first week of the workshop).

In his talk, Niels Obers reviewed an effective-field-theory-motivated
approach that captures a regime of the dynamics of higher-dimensional
black holes that has no counterpart in four dimensions. In this
so-called `blackfold' approach, black holes are regarded as black branes
whose worldvolume spans a curved submanifold of a background spacetime.
Obers also discussed how this method answers (in the affirmative) the
question of whether five-dimensional black hole spacetimes with only one
spatial Killing vector exist, mentioning such exotic solutions as
``helical black rings" (otherwise known as {\it slinkies}).

Mukund Rangamani gave a review of his work with Bhattacharyya, Hubeny
and Minwalla, where they show that the Einstein equations that govern
the dynamics of black branes in anti-deSitter space can be recast, for
perturbations of wavelengths much longer than the thermal and curvature
scale, in the form of fluid equations for a relativistic conformal
fluid. The properties of this fluid, including its shear viscosity as
well as other transport coefficients, can be systematically computed.
Although it fits naturally within the AdS/CFT correspondence (and its
possible application to fluids such as the quark-gluon plasma), the
result can be obtained and understood within a purely
general-relativistic context.

In a similar context, Oscar Dias reviewed his work (with Caldarelli,
Cardoso, Emparan, Gualtieri, and Klemm), which elevates the analogy
between black holes and lumps of fluid to a precise duality. This allows
comparison between the rich variety of phases of rotating fluid and
corresponding black hole solutions, and the instabilities that beset
them.

The AdS/CFT correspondence was also the context of the talk by Simon
Ross on holography for non-relativistic field theories, which reviewed
the behaviour of two separate classes of duals: the Schrodinger
spacetime which encodes field theories with Schrodinger symmetry
including Galilean boosts and the Lifshitz spacetime exhibiting
anisotropic scale invariance. An ongoing endeavor is to use these
gravity duals to define the stress tensor for the corresponding
non-relativistic field theories.

Through specific applications of the AdS/CFT correspondence, gravity has
started playing an increasingly prominent role in condensed matter
physics. This new field, dubbed AdS/CMT correspondence was the subject
of two talks, by Gary Horowitz and Andrei Parnachev. Gary Horowitz gave
a review of his work with Hartnoll and Herzog and further work with
Roberts on holographic superconductors. Remarkably, the gravitational
dual of a superconductor can be obtained in a Maxwell - (charged) scalar
- (AdS) gravity system, where charged hairy black holes are possible
below a certain critical temperature. Andrei Parnachev talked about
signatures of Fermi liquid formation in the dual field theory, using
D7-brane dynamics on the AdS background.

Another instance of this conceptual cross-breeding between field theory
and General Relavity was presented by Barak Kol. His approach, which
shares many of the concepts discussed in Obers' talk, takes advantage of
the well-developed ideas and techniques of effective quantum field
theory in order to efficiently solve problems of black hole collisions
and of `caged' black holes in Kaluza-Klein theories.

Gary Howowitz, joined by Monica Guica and Oscar Dias, led a discussion
on the Kerr/CFT correspondence proposed by Strominger~et~al.\ to relate
the near-horizon limit of the extremal Kerr black hole to a chiral
two-dimensional conformal field theory. Although this proposal correctly
accounts for the black hole entropy, the near-horizon (NHEK) geometry
admits no excited states, so the meaning of the correspondence is
presently unclear past the kinematic level.

Iosif Bena talked about black hole microstates and the information
paradox, reviewing the `fuzzball' proposal of Mathur, and describing the
construction of the candidate microstate geometries, which approach
certain extremal black holes outside the horizon but differ near and
inside the horizon. The lesson expounded by this programme, that
extremal black holes should be thought of as an ensemble of microstates
described by horizon-free and singularity-free `geometries', suggests
that quantum effects can modify classical physics even at long scales.

All through the duration of the workshop the Center was humming with the
discussions and intense interaction among participants (in spite of the
temptation to enjoy the good weather outdoors), and the overall open and
collaborative atmosphere was very much appreciated by everyone. It was
generally felt that such a successful and enjoyable meeting should have
a continuation in the future, and plans to hold a second Benasque
Workshop on Gravity in July 2011 are already underway.
\vfill\eject

\section*{\centerline
{Workshop on the Fluid-Gravity Correspondence}}
\addtocontents{toc}{\protect\medskip}
\addcontentsline{toc}{subsubsection}{
\it Workshop on the Fluid-Gravity Correspondence, 
by Mukund Rangamani }
\parskip=3pt
\begin{center}
Mukund Rangamani, Durham University 
\htmladdnormallink{mukund.rangamani-at-durham.ac.uk}
{mailto:mukund.rangamani@durham.ac.uk}
\end{center}

The workshop on the Fluid-Gravity correspondence was held at the Arnold Sommerfeld Center for theoretical physics which is part of the Ludwig-Maximillians-Universitat and is located in Munich, Germany. The main aim of the workshop which took place between September 2, 2009 and September 7, 2009, was  to bring together experts interested in the various aspects of the fluid-gravity correspondence, to have them report on recent developments and to induce an exchange of  ideas about the subject.

The fluid-gravity correspondence which itself is a consequence of the AdS/CFT or the gauge-gravity correspondence, relates the dynamics of  classical conformal relativistic hydrodynamics to the dynamics of classical general relativity in asymptotically Anti deSitter spacetimes. In particular, given a solution to the relativistic fluid equations of motion, there is a systematic procedure to construct an inhomogeneous, dynamical, asymptotically Anti deSitter black hole geometry, order by order in a particular perturbation expansion. 

The conference began on September 2, 2009 with a review talk by Andrei Starinets, who is one of the pioneers in the subject of using AdS/CFT to extract transport coefficients of strongly coupled plasmas. Andrei reviewed the traditional methods to compute transport coefficients using linear response theory (the famous Kubo formulae) in field theory and detailed how this may be done using the gravitational description via computing the retarded Green's function in black hole backgrounds. This was the basic set of calculations which has led to a lot of excitement in the subject owing to the famous bound of Kovtun, Son and Starinets (KSS) which asserts that the ratio of shear viscosity to entropy density of a relativistic system is bounded from below by $1/4\pi$ (in natural units). 

The afternoon talks were by Ram Brustein who discussed various aspects of the KSS bound. Since the bound was originally derived in two-derivative theories of gravity,  the talk explored higher derivative corrections, where it is well known that the bound can be violated (for instance by Gauss-Bonet couplings). This was followed by Yaron Oz, who discussed how the membrane paradigm picture of Damour, Price and Thorne can be realized in the fluid-gravity context.  Oscar Dias then spoke about his work on using the fluid description to understand instabilities of black hole spacetimes. In particular, he developed the nice analogy between Rayliegh-Plateau type instabilities of droplets (spinning and non-spinning) to the Gregory-Laflamme like instabilities of black holes and black strings and how this can be made more than an analogy in certain contexts within the AdS/CFT correspondence.

 The second morning was devoted to a review of the fluid-gravity correspondence itself. Shiraz Minwalla gave a fantastic talk on the basic framework that detailed how one relates fluid dynamics to gravity. This was followed by R. Loganayagam, who explained how the framework of the fluid-gravity correspondence can be used to understand aspects of known stationary solutions such as the rotating AdS black holes which take extremely simple forms when expressed in variables that are natural for fluid dynamics. The second afternoon began with my talk on using the AdS/CFT correspondence to understand the details of Hawking radiation of strongly coupled quanta and the existence of new black hole spacetimes. This was followed by Michal Heller who described the work of constructing holographic duals to the boost invariant fluid flow. The last talk of the day was by Veronika Hubeny who spoke about identifying the geometric dual of fluid entropy, pointing out that generically in dynamical systems one should not expect the event horizon to be associated with a sensible notion of entropy.
 
 The third day began with a very interesting talk by K. Sreenivasan on vortex dynamics in superfluids, which was followed by Alex Buchel reviewing recent developments on higher derivative corrections to transport coefficients. Sean Hartnoll then spoke about applications of AdS/CFT to condensed matter systems; one of the most intriguing aspects about the talk was the new formulation of the one-loop determinant about a Euclidean quantum gravity saddle point in terms of sum over quasi-normal mode spectrum. This was followed by Jan de Boer who explained how Hawking radiation on the world-sheet theory of strings which probe black hole backgrounds can be used to understand Brownian motion. 
The last talk of the day was by Harvey Reall who explained to us the Kerr-CFT correspondence which identifies the dynamics of the near horizon geometry of extremal Kerr (NHEK) geometry in four dimensions with a two dimensional chiral CFT. Harvey focussed on the gravitational aspects of the NHEK geometry and demonstrated that there is no dynamics in this spacetime. In fact, there is no initial data which preserves the constraints under evolution. 

The fourth day was devoted to applications of AdS/CFT to condensed matter physics. We heard from Pavel Kovtun, Chris Herzog and Andreas Karch the recent developments in identifying gravity duals for non-relativistic systems, aspects of superfluids etc.. Superfluidity is seen in a very simple model of gravity in AdS coupled to a Abelian Higgs model. Above a certain critical temperature one finds that the Reissner-Nordstorm AdS black holes dominate the grand canonical potential, while below one encounters novel scalar hair black holes. We heard how these solutions can be constructed numerically and probed using various techniques, and how the physics matches quite nicely with that seen in real superfluids. 

The last day of the conference began with a review by David Mateos on using the AdS/CFT correspondence to understand properties of the quark-gluon plasma, followed by a few talks again on holographic superconductors. Johanna Erdmenger reviewed the superconducting state in Einstein Yang-Mills theory and Matthias Kaminski described a new efficient way to compute quasi-normal modes for theories with non-trivial interactions between the matter and gravitational sector. Silviu Pufu decribed how the holographic superconducting/superfluid instability can be realized in string theory and the last talk of the conference was by Michael Lublinsky.

The conference dinner took place on September 4, 2009 at Geogenhof restaurant a few blocks away from the Arnold Sommerfeld Center; this provided a great opportunity for the participants to continue the discussions on the questions raised by the various talks. On Saturday, since the afternoon session was free, Martin Ammon, a local student organized a guided tour around the center of Munich.

\end{document}